\pdfoutput=1

\documentclass[11pt]{article}

\usepackage{authblk}
\usepackage[final]{acl}
\usepackage{bm}
\usepackage{booktabs}
\usepackage{times}
\usepackage{latexsym}
\usepackage{amssymb}
\usepackage{float}
\usepackage{subcaption}
\usepackage{caption}
\usepackage{arydshln}
\usepackage{verbatimbox}
\usepackage[T1]{fontenc}

\usepackage[utf8]{inputenc}

\usepackage{microtype}
\usepackage{graphicx}
\usepackage{inconsolata}
\usepackage {multirow}
\usepackage{amsmath}

\title{Voice Attribute Editing with Text Prompt}

\author[1]{\textbf{Zhengyan Sheng}}
\author[1]{\textbf{Yang Ai}}
\author[2]{\textbf{Li-Juan Liu}}
\author[2]{\textbf{Jia Pan}}
\author[1 \thanks{Corresponding author}]{\textbf{Zhen-Hua Ling}}
\affil[1]{National Engineering Research Center of Speech and Language Information Processing, \protect\\ University of Science and Technology of China, Hefei, China}
\affil[2]{iFLYTEK Research, Hefei, China}
\affil[ ]{\texttt{zysheng@mail.ustc.edu.cn, \{yangai, zhling\}@ustc.edu.cn}, \\ \{ljliu, jiapan\}@iflytek.com}

\begin{document}
\maketitle
\begin{abstract} 
Despite recent advancements in speech generation with text prompt providing control over speech style, voice attributes in synthesized speech remain elusive and challenging to control. This paper introduces a novel task: voice attribute editing with text prompt, with the goal of making relative modifications to voice attributes according to the actions described in the text prompt.
To solve this task, \textit{VoxEditor}, an end-to-end generative model, is proposed. In \textit{VoxEditor}, addressing the insufficiency of text prompt, a Residual Memory (ResMem) block is designed, that efficiently maps voice attributes and these descriptors into the shared feature space. Additionally, the ResMem block is enhanced with a voice attribute degree prediction (VADP) block to align voice attributes with corresponding descriptors, addressing the imprecision of text prompt caused by non-quantitative descriptions of voice attributes.
We also establish the open-source \textit{VCTK-RVA} dataset, which leads the way in manual annotations detailing voice characteristic differences among different speakers.
Extensive experiments demonstrate the effectiveness and generalizability of our proposed method in terms of both objective and subjective metrics. The dataset and audio samples are available on the website \footnote{
\url{https://anonymous.4open.science/w/VoxEditor_demo-ACL/}}.
\end{abstract}

\section{Introduction}
Voice characteristics, serving as an expression of the speaker's identity, is a crucial component of speech.  Effectively controlling voice characteristics in speech has consistently been a focal point in research.
Voice conversion (VC) \cite{mohammadi2017overview} stands out as a representative technology that seeks to change the voice characteristics from a source speaker to a target speaker while preserving the linguistic content. Traditional VC tasks depend on reference audio, but finding suitable reference audio is always challenging, especially for the applications like personalized voice creation for virtual characters and automatic movie dubbing. Given that natural language acts as a convenient interface for users to express the voice attributes, which refer to human perception of voice characteristics (e.g., "husky", "bright", "magnetic"), using text prompts \cite{guo2023prompttts, ji2023textrolspeech} as guidance is a more viable approach to flexible voice creation.
\captionsetup{skip=5pt}

\begin{figure}[!t]
\centering
\includegraphics[width=2.5in]{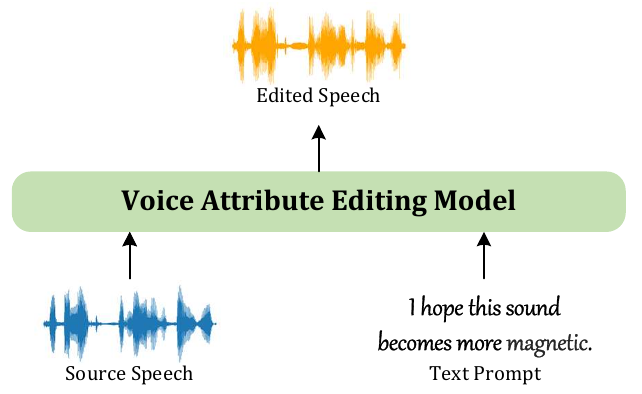}
\caption{Illustration of voice attribute editing with text prompt.}

\label{fig_1}
\vspace{-0.5cm}
\end{figure}

This paper introduces a novel task: voice attribute editing with text prompt. 
As shown in Figure \ref{fig_1}, given source speech and a text prompt that describes the desired editing actions, i.e., relative modifications on specific voice attributes,  the task aims to alter the source speech according to the text prompt while keeping the linguistic content unchanged. 
The voice attribute editing task fundamentally differs from recent speech generation tasks with text prompt \cite{ji2023textrolspeech, watanabe2023coco, yang2023instructtts}. These related tasks utilize text prompt for voice control rather than reference audio, resulting in speech style (e.g., gender, emotion and rhythm) that roughly matches the input text prompt, but they lack the ability to finely modify specific voice attributes. Specific distinctions are outlined in Section \ref{Section2.1}.



The primary two challenges encountered in the voice attribute editing task are the \textbf{insufficiency} and \textbf{imprecision} of text prompt. First, the insufficiency refers to the challenge posed by the multi-dimensional nature of the voice perception space, making it difficult for text prompt to fully capture all voice characteristics. This difficulty is exacerbated in the voice attribute editing task that only modifies specific voice attributes, thus further complicating the establishment of mapping from the text prompt to corresponding voice attributes.
Second, the imprecision means that we always express our perception of voice characteristics through qualitative descriptors rather than relying on quantitative physical descriptors \cite{wallmark2018describing}. For the voice attribute editing task, 
this results in ambiguity when expressing the detailed differences in particular voice attributes between speakers. 

To address the aforementioned challenges, we propose \textit{VoxEditor}, the first model to deal with the voice attribute editing task.  
To tackle the insufficiency issue, we propose a Residual Memory (ResMem)-based text-voice alignment module.  The ResMem consists of two components: the main memory,  which  utilizes trainable slots to quantize the common space for text and voice characteristics, and the residual memory, which compensates for challenging-to-describe aspects in voice characteristics. To address the imprecision issue,  the ResMem block is enhanced with the voice attribute degree prediction (VADP) module, designed to predict the difference degree of the specific voice attribute between two speakers. During inference, with the assistance of a large language model (LLM), we first extract voice attribute descriptors from the text prompt. Subsequently, we perform semantically meaningful interpolation between the descriptor embedding and the source speaker embedding, resulting in the edited speaker embedding for generating speech.  

To facilitate research on the voice attribute editing task, this paper presents a manually annotated dataset, \textit{VCTK-RVA}, which annotates Relative Voice Attribute differences between same-gender speakers based on the VCTK \cite{vctk} corpus. 
Initially, speech experts distilled a descriptor set from a large-scale internal speech dataset to describe common voice attributes. 
Then, speech experts conducted pairwise comparisons of voice characteristics among same-gender speakers in the open-source VCTK corpus and selected suitable descriptors from the set to effectively capture the major differences in voice characteristics. 


To validate the effectiveness and generalizability of our method, we meticulously devised several metrics for the task. Experimental results show that \textit{VoxEditor} can generate high-quality speech that align well with the input text prompt and preserve voice characteristics of the source speech to some extent.  These results highlight  the controllability, generality, and
quality of \textit{VoxEditor}.

We summarize our main contributions as follows. Firstly, we introduce a new task: voice attribute editing with text prompt. This task enables users to make relative modifications to voice attributes in source speech based on the provided text prompt, offering a convenient method for creating desired voice characteristics. 
 Secondly, we construct the \textit{VCTK-RVA} dataset, an open-source resource that pioneers in describing differences in voice characteristics between speakers. Thirdly, \textit{VoxEditor} is proposed for the VE task, which integrates ResMem and VADP modules to overcome challenges caused by the insufficiency and imprecision of text prompt.

\section{Related Work}
\subsection{Speech Generation with Text Prompt}
\label{Section2.1}
Considering the success of text-guided generation in both text and images, many recent works have explored speech generation with text prompt \cite{guo2023prompttts, leng2023prompttts, ji2023textrolspeech}. However, only a few of these works focus on specific aspects of voice characteristics \cite{shimizu2023prompttts++, zhang2023promptspeaker, yao2023promptvc}, primarily addressing speech style factors such as gender, speaking speed, energy, and emotion. 

In the context of methods, these studies commonly incorporated BERT\cite{kenton2019bert} to extract textual embeddings from the input text prompt and utilized supervised training with reference speech to establish a mapping from textual embeddings to speaker embeddings. 
Prompttts2 \cite{leng2023prompttts} introduced Diffusion \cite{ho2020denoising} sampling to mitigate the insufficiency of text prompt. Nevertheless, the diversity achieved during inference remains both elusive and beyond control.
The proposed \textit{VoxEditor} employs the ResMem and VADP blocks to address the insufficiency and imprecision issues of text prompt, allowing users to relatively modify the target voice attribute.

Several speech datasets \cite{ji2023textrolspeech, watanabe2023coco, yang2023instructtts} enriched with text prompt have also been established. These datasets provided the individual text descriptions of speech style for each speech sample,  which failed to convey the detailed differences in voice attributes between speakers and were unsuitable for the voice attribute editing task. In contrast, relative attribute annotations in the \textit{VCTK-RVA} dataset allow the speech to be ranked across various voice attributes,  facilitating alignment between independent voice attributes and the corresponding descriptors. 

\begin{table}[]
\centering
\begin{tabular}{lc lc}
\toprule[1pt]
\textbf{Descriptor} & \textbf{Freq.} & \textbf{Descriptor} & \textbf{Freq.} \\ 
\cmidrule(lr){1-2} \cmidrule(lr){3-4}
Bright      & 17.10      & Thin        & 13.03      \\
Coarse      & 11.62      & Slim        & 11.31      \\
Low         & 7.43       & Pure        & 5.48       \\
Rich        & 4.71       & Magnetic    & 3.64       \\
Muddy       & 3.59       & Hoarse      & 3.32       \\
Round       & 2.48       & Flat        & 2.15       \\
Shrill      & 2.08       & Shriveled   & 1.74       \\
Muffled     & 1.44       & Soft        & 0.82       \\
Transparent & 0.66       & Husky       & 0.59       \\ 
\bottomrule[1pt]
\end{tabular}
\caption{The descriptor set is used for describing the common voice attributes, and the Freq. represent frequency (\%) of each descriptor in \textit{VCTK-RVA}.}
\label{table1}
\vspace{-0.5cm}
\end{table}

\subsection{Memory Network}
Memory Network \cite{weston2015memory} incorporates a long-term memory module with the ability to be both read from and written to. 
Recently, Key-value memory has been employed for cross-modal alignment across various tasks \cite{chen2021cross, sheng2023face}.
However, these methods often  neglect the information gap between different modalities. Given the insufficiency issue of text prompt, we propose the ResMem block to bridge the gap between the text prompt and voice characteristics. 

\section{\textit{VCTK-RVA} Dataset}

\subsection{Descriptors for Voice Attributes} 
\label{descripitor set}
To construct a dataset suitable for the voice attribute editing task,  manual annotations are necessary to express voice perception. However, systematic research on the voice perception space is currently lacking. Therefore, for practicality, we adopt a descriptor set to describe the differences in voice characteristics among speakers. Here, the descriptor set refers to the keywords commonly used in natural language to describe voice characteristics.
In terms of engineering implementation, the descriptor set should be concise and cover the most commonly used voice characteristic descriptions.

Specifically, we engaged 10 speech experts with professional backgrounds in acoustic research to describe  the voice characteristics of 1500 speakers based on internal 26-hour recordings in natural language. Then we merged synonyms of keywords in these descriptions and summarized the descriptor set based on word frequency statistics, as shown in Table \ref{table1}.

\begin{figure*}[!t]
\centering
\includegraphics[width=6.2in]{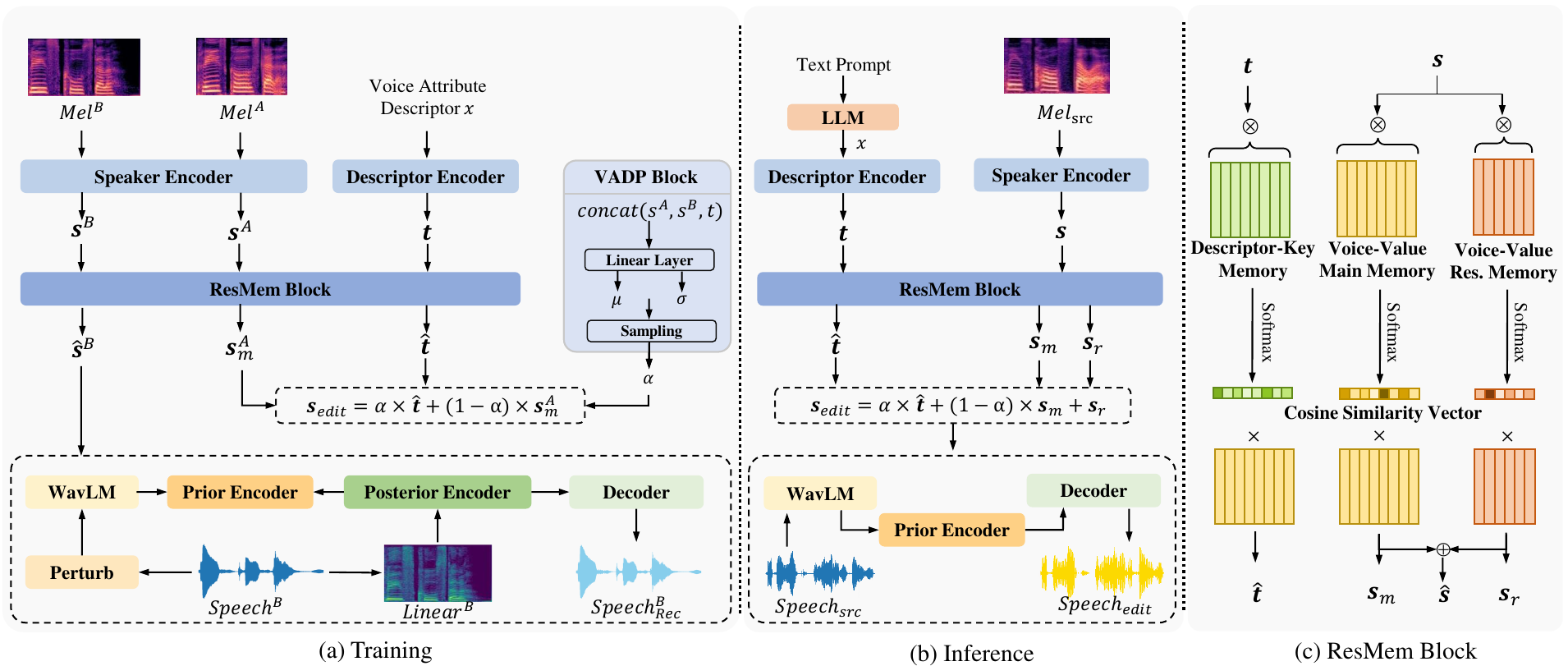}
\caption{The overall flowchart of our proposed \textit{VoxEditor}. During the training process,
two speech segments ($Speech^A$ and $Speech^B$) are used, along with voice attribute descriptor $x$. In the inference process, the model takes source speech and the text prompt as inputs to generate edited speech. Here $Mel$ denotes the Mel spectrograms, $Linear$ denotes the linear spectrograms, $\bm{s}_m$ demotes the recalled main speaker embeddings, $\bm{s}_r$ denotes the recalled residual speaker embeddings and $\hat{\bm{s}}$ denotes the recalled speaker embeddings.}
\label{fig_2}
\vspace{-0.5cm}
\end{figure*}


\subsection{Voice Attribute Annotations}
We choose the publicly available VCTK (version 0.92) dataset \cite{vctk}, which has been widely utilized in VC and text-to-speech studies, as the materials for annotation. The VCTK dataset consists of speech sentences of 110 speakers, including  62 females and 48 males. Each speaker utters approximately 400 sentences in a reading style, resulting in a total of 43,475 sentences. 

We hired four speech experts to conduct pairwise comparisons of voice characteristics of same-gender speakers. When presented with speech samples from $SpeakerA$ and $SpeakerB$,  speech experts  listened to the samples to identify the differences in voice attributes between these two speakers. Subsequently, these experts selected an unrestricted number of descriptors $\bm{v}$ from the descriptor set built in Section \ref{descripitor set} to express that $SpeakerB$ exhibits more prominent voice attributes $\bm{v}$ when compared to $SpeakerA$. This forms an annotated tuple,
$\{SpeakerA,  SpeakerB,  \bm{v}\}$, where $ \bm{v}  \in D \cup \text{"Similar"}$, with  $D$ representing  the descriptor set and 
"Similar" indicating  that speech experts perceived the voice characteristics of two speakers as highly similar.  
In cases of annotation discrepancies, experts engaged in discussions to reach a final consensus. 
The current annotation only considers a one-way form, highlighting the more prominent voice attributes and not annotating the diminished  voice attributes. Overall, through same-gender pairwise comparisons among 62 male and 48 female speakers, we collected a total of $62 \times 61 + 48 \times 47 = 6038$ annotated data points. In the entire dataset, the percentages of $\bm{v}$ corresponding to one, two, and three descriptors are 71.19\%, 26.84\%, and 1.97\%, respectively. The "Similar" label accounts for 6.81 \%.

We randomly selected 200 annotated samples and uploaded them on Amazon Mechanical Turk (AMT), inviting ordinary individuals to evaluate the annotations. Specifically, we provided multiple speech samples from two speakers along with our voice attribute annotations. Listeners were instructed to listen to the speech using headphones and determine whether they agreed with our annotations. A total of 40 listeners from AMT participated in the test, and their average agreement rate reached an impressive 91.78 \%.

\section{VoxEditor}

\subsection{Overall Architecture}
Our proposed \textit{VoxEditor} adopts the end-to-end auto-encoder paradigm \cite{li2023freevc}, that decomposes speech into content and speaker embeddings, subsequently re-synthesizing the content and  edited speaker embeddings into new speech.  
As shown in Figure \ref{fig_2} (b), during inference, the LLM initially analyzes the input text prompt to obtain a specific descriptor indicating the desired  voice attribute for modification. Subsequently, the descriptor embedding and speaker embedding are extracted from the descriptor and the Mel spectrograms of the source speech using a descriptor encoder and a speaker encoder. These embeddings are then input into the ResMem block (described in Section \ref{ResMem Block}), and the output is combined through linear interpolation to derive the edited speaker embeddings. Simultaneously, the pretrained WavLM \cite{chen2022wavlm} and prior encoder are employed to extract the content embedding from source speech. Finally, the content and speaker embeddings are fed into the decoder, which generates the edited speech. It is worth noting that when the text prompt contains multiple voice attribute descriptors, the source speech needs to be edited sequentially based on these descriptors.

During training, \textit{VoxEditor} is provided with an annotated tuple $\{SpeakerA, SpeakerB, \bm{v}\}$. As shown in Figure \ref{fig_2} (a), it takes speech segments $Speech^A$ and $Speech^B$, along with a randomly selected voice attribute descriptor $x \in \bm{v}$, as input. These speech segments are randomly segmented from the respective speaker's speech.
In addition to the mentioned modules, the VADP block (described in Section \ref{VADP Block}) is employed to predict the difference degree of the specific voice attribute between two speakers. Other model modules and the autoencoder training strategy follow FreeVC \cite{li2023freevc}, which adopts variational inference augmented with a normalizing  flow and an adversarial training process. Further details about the perturb-based data augmentation, prior encoder, posterior encoder and pretraining strategy can be found in Appendix \ref{appendix A}. 

\subsection{ResMem Block} 
\label{ResMem Block}
Considering the insufficiency issue of text prompt, we employ a ResMem block to establish a mapping between voice attributes and their corresponding descriptors within the same feature space. Illustrated in Figure \ref{fig_2}(c), 
the ResMem module accepts either the speaker embedding ${\bm{s}\in \mathbb{R} ^D}$ or the descriptor embedding ${\bm{t}\in \mathbb{R} ^D}$ as input, where ${\bm{s}}$ and ${\bm{t}}$ are derived from the pretrained speaker encoder and the descriptor encoder, respectively, and $D$ represents the dimension of descriptor or speech embeddings. 
The ResMem block is composed of a main voice-value memory $\bm{M}_{mv} \in \mathbb{R}^{M \times D}$,
a residual voice-value memory $\bm{M}_{rv} \in \mathbb{R}^{N \times D}$
and a descriptor-key memory ${\bm{M}_{k}\in \mathbb{R} ^{M \times D}}$,
where $M$ denotes the number of the slots in ${\bm{M}_{mv}}$ and ${\bm{M}_{k}}$,  $N$ denotes the number of slots in ${\bm{M}_{rv}}$ and $D$ is the dimension for each slot, which equals to the dimension of the descriptor and speaker embeddings. 

${\bm{M}_{mv}}$ is designed to capture the main voice characteristics that can be articulated with voice descriptors, while ${\bm{M}_{rv}}$ is utilized to address aspects of voice characteristics that are challenging to describe in text. Specifically, when a speaker embedding ${\bm{s}}$ is given as a query, the cosine similarity between the query and each slot in ${\bm{M}_{mv}}$ is computed, followed by softmax normalization function, expressed as follows,
\setlength{\abovedisplayskip}{8pt}
\setlength{\belowdisplayskip}{10pt}
\begin{equation}
w_{mv}^i=softmax(\frac{ \bm{s}^\top \bm{m}_{mv}^i}{\left\|\bm{s}\right\|_2\left\|\bm{m}_{mv}^i\right\|_2 }),
\end{equation}
where $\bm{m}_{mv}^i$ denotes the $i$-th slot in ${\bm{M}_{mv}}$ and $w_{mv}^i$ represents the degree of relevance between the $\bm{m}_{mv}^i$ and the speaker embedding ${\bm{s}}$. We then obtain the cosine similarity vector ${\bm{w}_{mv}}=[w_{mv}^1, w_{mv}^2, \cdots, w_{mv}^M] ^\top \in \mathbb{R} ^M$ by computing cosine similarity with all slots. Next, we generate the recalled main speaker embedding as follows,
\begin{equation}
\hat{\bm{s}}_{m}= \bm{M}_{mv}^\top \bm{w}_{mv}.
\end{equation}
Similarly, we generate the recalled residual speaker embedding as follows,
\begin{equation}
w_{rv}^j=softmax(\frac{ \bm{s}^\top \bm{m}_{rv}^j}{\left\|\bm{s}\right\|_2\left\|\bm{m}_{rv}^j\right\|_2 }),
\end{equation}
\begin{equation}
\bm{w}_{rv}=[w_{rv}^1, w_{rv}^2, \cdots, w_{rv}^N] ^\top \in \mathbb{R} ^N,
\end{equation}
\begin{equation}
\hat{\bm{s}}_{r}= \bm{M}_{rv}^\top \bm{w}_{rv},
\end{equation}
where $\bm{m}_{rv}^j$ denotes the $j$-th slot in ${\bm{M}_{rv}}$.
Then, we obtain the recalled speaker embedding $\hat{\bm{s}}$ and calculate the mean square error (MSE) between  $\bm{s}$ and $\hat{\bm{s}}$ as well as the MSE between  $\bm{s}$ and $\hat{\bm{s}}_{m}$.
\begin{equation}\label{eq_6}
\hat{\bm{s}}= \hat{\bm{s}}_{m} + \hat{\bm{s}}_{r},
\end{equation}
\begin{equation}
\mathcal{L}_{rec}=\| \bm{s} - \hat{\bm{s}} \Vert _2 ^2 + \| \bm{s} - \hat{\bm{s}}_{m} \Vert _2 ^2.
\label{eq7}
\end{equation}
In this manner, the slots within the ${\bm{M}_{mv}}$ and ${\bm{M}_{rv}}$ can serve as foundational vectors for constructing the entire voice characteristics space, allowing for various combinations of these slots to represent a wide range of voices.

Then, we utilize the slots in $\bm{M}_{mv}$ as a streamlined bridge to map descriptor embeddings onto the voice space. In detail, given the descriptor embedding $\bm{t}$, we generate the recalled descriptor embedding $\hat{\bm{t}}$ in a similar way as the recalled main speaker embedding as follows,
\begin{equation}
w_{t}^i=softmax(\frac{ \bm{t}^\top \bm{m}_{k}^i}{\left\|\bm{t}\right\|_2\left\|\bm{m}_{k}^i\right\|_2 }),
\end{equation}
\begin{equation}
\bm{w}_{t}=[w_{t}^1, w_{t}^2, \cdots, w_{t}^M]^\top \in \mathbb{R} ^M,
\end{equation}
\begin{equation}
\hat{\bm{t}} = \bm{M}_{mv}^\top \bm{w}_{t},
\end{equation}
 where $m_k^i$ denotes the $i$-th slot in $\bm{M}_k$, cosine similarity is calculated with the slots in the descriptor-key memory ${\bm{M}_{k}}$ and aligned with the main voice-value memory ${\bm{M}_{mv}}$. In this way, descriptor embeddings are mapped to the main voice characteristics space, with the slots in $\bm{M}_{mv}$ serving as the basis vectors.

 

\subsection{VADP Block} 
\label{VADP Block}
Considering the imprecision inherent in text prompt, we propose the VADP block, designed to predict the difference degree of the specific voice attribute between two speakers. Voice characteristics can exhibit local variations \cite{zhou2022content} due to factors such as content, rhythm, and emotion. Therefore, we assume that the differences in voice attributes between different speech samples from two speakers are not deterministic but follows a Gaussian distribution.  

 As shown in Figure \ref{fig_2}(a), given the speaker embedding ${\bm{s}^A}$ from $Speech^A$, speaker embedding ${\bm{s}^B}$ from $Speech^B$ and descriptor embedding $t$, we concatenate three embeddings to obtain a cross-modal embedding, We then use linear layers and ReLU activation functions to predict the mean and variance of a Gaussian distribution. Subsequently, we sample from this Gaussian distribution to obtain the difference degree of specific voice attributes, which we map through a sigmoid activation function to derive the editing degree $\alpha \in [0, 1]$. Similar to Imagic \cite{kawar2023imagic}, we linearly interpolate between the recalled descriptor embedding $\hat{\bm{t}}$ and the recalled main speaker embedding $\hat{\bm{s}}_{m}^A$ from $Speech^A$ to derive the edited speaker embedding ${\bm{s}_{edit}}$,
 \begin{equation}
{\bm{s}_{edit}} = \alpha \cdot \hat{\bm{t}} + (1-\alpha) \cdot \hat{\bm{s}}_{m}^A.
\label{eq11}
\end{equation}
 Additionally, we align the slot-weights distributions between the recalled main speaker embedding $\hat{\bm{s}}_{m}^B$ and ${\bm{s}_{edit}}$ using Kullback-Leibler (KL) divergence, 
\begin{equation} 
\mathcal{L}_{align}=D_{KL}(\bm{w}_{mv}^{B}||\alpha \cdot \bm{w}_t + (1-\alpha) \cdot \bm{w}_{mv}^{A}),
\label{eq12}
\end{equation}
where $\bm{w}_{mv}^{A}$ and $\bm{w}_{mv}^{B}$ denote the cosine similarity vectors in the main voice-value memory ${\bm{M}_{mv}}$ for $Speech^A$ and $Speech^B$, respectively. In this way, we explicitly align the voice attributes with their corresponding descriptors.

\subsection{Speaker Embedding Editing}
As shown in Figure. \ref{fig_2}(b),
we utilize a LLM (GPT-3.5-TURBO) to scrutinize the input text prompt and extract target voice attribute descriptors. In cases where the descriptor is absent from the set, the LLM is employed to locate the closest matching descriptor within the set for substitution. For further elucidation, please refer to Appendix \ref{appendix B}. Next, we input both the target voice attribute descriptor and source speech to obtain the recalled descriptor embedding $\hat{\bm{t}}$, recalled main  speaker embedding $\hat{\bm{s}}_{m}$ and recalled residual  speaker embedding $\hat{\bm{s}}_{r}$. Subsequently, we achieve edited speaker embedding through linear interpolation,
\begin{equation}
{\bm{s}_{edit}} = \alpha \cdot \hat{\bm{t}} + (1-\alpha) \cdot \hat{\bm{s}}_{m} + \hat{\bm{s}}_{r},
\label{eq13}
\end{equation}
where the value of editing degree  $\alpha$ is initially set to its recommended value 0.7 (refer to Figure 5). The editing degree can be further adjusted within the range $[0, 1]$, and increasing $\alpha$ will progressively enhance the prominence of the target voice attribute.

\section{Experiments}
\subsection{Implementation Details}
Our experiments were conducted using the \textit{VCTK-RVA} dataset, consisting of  98 speakers for both the training and validation sets. Among these, 200 sentences were randomly selected for validation, while the remaining sentences were utilized for training. For testing, speech samples were categorized into two sets: the seen speaker set, comprising speakers from the training set, and the unseen speaker set, consisting of speakers not encountered during training. Each set comprised 12 speakers, each contributing three sentences. 
During the evaluation, source speech underwent voice attribute editing for each voice attribute in the descriptor set, with $\alpha$ varying from 0 to 1 in increments of 0.1. We devised 10 predefined sentence patterns, such as "\textit{I want this sound to become more [Descriptor]}". For each edit, a sentence pattern was randomly chosen, and \textit{[Descriptor]} was replaced with the target voice attribute descriptor to form the text prompt. 

All speech samples were downsampled to 16k Hz.
Linear spectrograms and 80-band Mel spectrograms were computed using a short-time Fourier transform (STFT) with FFT, window, and hop sizes set to 1280, 1280, and 320, respectively. The dimensions of descriptor embeddings, speaker embeddings and slots were equal to $D=256$. The numbers of slots in the ResMem Block were set to $M = 32$ and $N = 4$. We put more information about $M$ and $N$ in Appendix \ref{App ResMem block}.

\begin{table*}[]

\resizebox{\textwidth}{!}{
\begin{tabular}{lcccccccc}

\toprule
\multicolumn{1}{c}{\multirow{2}{*}{Model}} & \multicolumn{4}{c}{Seen}     & \multicolumn{4}{c}{Unseen}   \\ 
\cmidrule(r){2-5} \cmidrule(l){6-9} \multicolumn{1}{c}{} 
                                          & TVAS             & MOS-Nat          & MOS-Cons        & MOS-Corr         & TVAS           & MOS-Nat          & MOS-Cons  & MOS-Corr \\ \hline
PromptStyle                                   & 0.0089           & 3.9200           & 2.0542          & 2.0321           & 0.0047         & 3.9112           & 1.9914    &2.5532\\
VoxEditor                                  & \textbf{0.0574} & 3.9147   & \textbf{3.5333} &\textbf{3.7100}  & \textbf{0.0561} &3.9077  & \textbf{3.4701}          &\textbf{3.7036} \\ \hline
w/o Voice Res.                            &0.0559           &3.9194             &  3.4942         &3.4739          &0.0553            &3.9069          & 3.4586          &3.4357\\
w/o ResMem                                 &0.0102           &3.8906             &  2.1142         &2.1934          &0.0098           &3.9073            & 2.3038          &2.6086\\
w/o VADP                                   &0.0526          &3.9146             &  3.4342         &3.6967          &0.0504            &3.9105           & 3.3176          &3.6786 \\ \bottomrule

\end{tabular}}
\caption{Objective and subjective evaluation results of comparison methods. The definitions of all metrics can be found in Section \ref{metrics}.}
\label{table_2}
\end{table*}

\begin{figure*}[!t]
\centering
\includegraphics[width=6.2in]{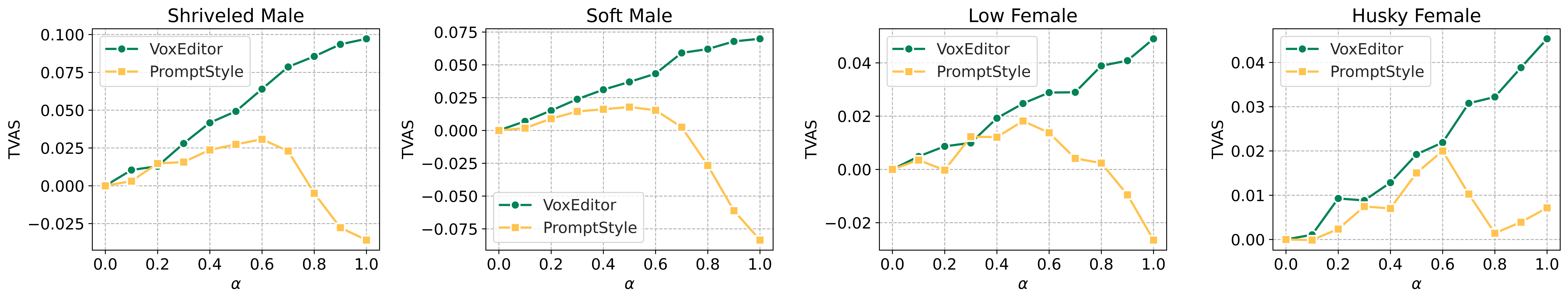}
\caption{The variation of the TAVS metric for generated speech edited with different attributes under various values of editing degree $\alpha$.}
\label{fig_3}
\vspace{-0.5cm}
\end{figure*}

\subsection{Metrics}
\label{metrics}
\paragraph{Objective Metrics} 

To objectively assess whether the edited voice characteristics align with the text prompt, we introduced Target Voice Attribute Similarity (TVAS) metrics. Since there were no target speakers, we statistically derived reference speakers for each gender corresponding to each voice attribute. 
Specifically, a speaker occurring as the $SpeakerB$ in an annotated tuple $\{SpeakerA,  SpeakerB,  \bm{v}\}$ was defined as one of the reference speakers for the descriptor $x$, where $x \in \bm{v}$.
We traversed the entire dataset to obtain the reference weight $\eta_{x}^{j}= \frac{occ_{j}}{number_{x}}$ of the $j$-th reference speaker of voice attribute $x$, where $j \in [1,k] $, $k$ was the number of reference speakers of the voice attribute $x$, ${occ_{j}}$ represented the occurrence number of the $j$-th reference speaker of voice attribute $x$ and $number_{x}$ was the total occurrence number of voice attribute $x$. A higher reference weight indicated a more prominent voice attribute of the reference speaker.
Additionally, we applied the well-known open-source speaker verification toolkit, WeaSpeaker\footnote{https://github.com/wenet-e2e/wespeaker}, to extract speaker embeddings of edited speech and obtain mean speaker embeddings of each reference speaker. When source speech was edited with target voice attribute $x$, we calculated the cosine similarity scores $o^j$ between the speaker embeddings of edited speech and the mean speaker embedding of the $j$-th reference speaker of the attribute $x$. Then all cosine similarity scores were weighted with corresponding reference weight, resulting in the Absolute Target Voice Attribute Similarity metrics under difference value of $\alpha$ ($\text{ATVAS}_{\alpha}= \sum_{j=1}^{j=k} o^j \cdot \eta_{x}^{j}$). 
Then, to better focus on the relative change of voice attribute similarities, we calculated the TVAS metrics with varying editing degree $\alpha$, $\text{TVAS}_{\alpha} = \text{ATVAS}_{\alpha} - \text{ATVAS}_{0}$, and averaged $\text{TVAS}_{\alpha}$ over all $\alpha$ to obtain the final TVAS metric for editing a source speech sample on the target voice attribute.
\paragraph{Subjective Metrics}
We employed three mean opinion scores to assess various aspects of the generated speech: speech naturalness (MOS-Nat), descriptor-voice consistency (MOS-Cons), and correlation between source and generated speech (MOS-Corr). MOS-Nat quantitatively measured the naturalness of the generated speech, with scores ranging from 1 (completely unnatural) to 5 (completely natural). MOS-Cons evaluated the consistency between the voice characteristics of the generated speech and the text prompt, with scores ranging from 1 (completely inconsistent) to 5 (completely consistent). Additionally, when the editing degree approaches 1, the generated speech should still retain some voice characteristics of the source speech. Therefore, MOS-Corr was introduced to assess the voice characteristics similarity between the generated speech and source speech when the editing degree approaches 1, with the score ranging from 1 (completely unrelated voice) to 5 (very similar voice). 100 generated speech samples covering each voice attribute were randomly selected for each subjective evaluation.  Three subjective metrics were evaluated on the AMT platform, and 20 listeners participated in the test each time.
\subsection{Evaluation Results}
\label{Evaluation Results}
As pioneers in addressing the voice attribute editing task with no existing comparable methods, we conducted a thorough comparative analysis of our proposed method against the following baseline and ablation models to evaluate its effectiveness: 
(1) PromptStyle \cite{liu2023promptstyle}: We utilized its style embedding generation module to replace the edited speaker embedding generation module in \textit{VoxEditor}. Specifically, the original prompt encoder was modified to a speaker encoder and a descriptor encoder.  MSE loss was employed to minimize the distance between $\bm{s}^{A}+\bm{t}$ and $\bm{s}^{B}$. (2) w/o Voice Res., (3) w/o ResMem, (4)w/o VADP. More details about these ablation models can be found in Appendix \ref{Ablation Models}. All objective and subjective evaluation results are summarized in Table. \ref{table_2}.

\begin{figure*}[!t]
\centering
\includegraphics[width=6.2in]{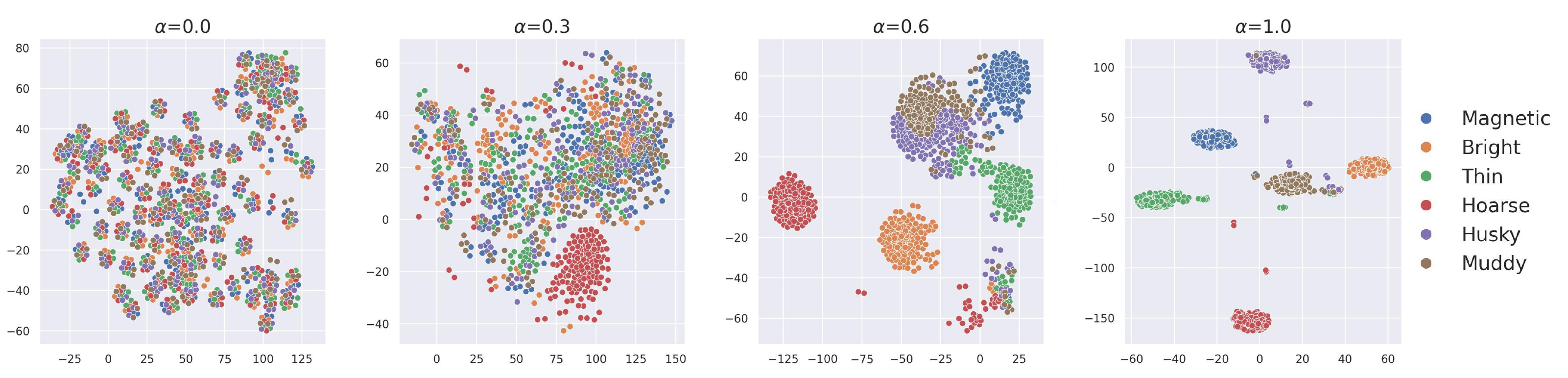}
\caption{The t-SNE visualization of the speaker embeddings extracted from generated speech edited with different attributes under various values of $\alpha$.}
\label{fig_4}
\vspace{-0.5cm}
\end{figure*}

We can observe that \textit{VoxEditor} outperformed other methods significantly in all metrics  (${p<0.05}$ in paired t-tests) except for MOS-Nat.  When the ResMem block was not utilized (PromptStyle and w/o ResMem), the model's performance sharply declined. This is attributed to the inability to effectively align the voice attributes with their corresponding descriptors. Furthermore, the TVAS metric for same-gender speakers with different editing degrees $\alpha$ is illustrated  in Figure \ref{fig_3}. We noticed  that as $\alpha$ increases, the speech generated by our method became increasingly prominent in the specified voice attributes. In contrast, for the PromptStyle method, the direction of editing was not consistent with the specified voice attributes.

We found that the performance of the w/o Voice Res. was comparable to that of our method in terms of TVAS and MOS-Cons, but there was a significant difference in MOS-Corr. For same-gender speakers, the voice characteristics of the generated speech by w/o Voice Res. were very similar when edited in the same voice attribute with $\alpha$ approaching 1, and the correlation in voice characteristics between the generated speech and the source speech was almost nonexistent.
The removal of the VADP block prevented the model from  modelling refined edited speaker embeddings during training, resulting in compromised descriptor embeddings. Consequently, during inference, even when the editing degree approached 1, the edited voice attributes were not prominent enough, leading to a decrease in the TVAS and MOS-Cons scores. In addition, since these methods follow the same auto-encoder paradigm, their performance in terms of MOS-Nat was quite comparable.



\subsection{Visual Analysis}
We randomly selected an additional set of 100 utterances from an unseen speaker and visualized the speaker embeddings of the generated speech edited with different voice attributes through t-SNE \cite{chan2019gpu}, as shown in Figure \ref{fig_4}. We observed that, as the editing degree $\alpha$ increased, generated speech edited with the same target voice attribute gradually 
formed distinct clusters, which demonstrated the stability of our method. Additionally, due to variations in the number of voice attribute annotations and differences in the prominence of voice attributes, there were variations in the editing performance of different voice attributes. When $\alpha$ equalled  0.3, the speech edited on the "Hoarse" attribute already exhibited clear voice features, forming clusters, while generated speech edited with other voice attributes predominantly retained the voice characteristics of the source speech.

\subsection{User Study}
\begin{figure}
  \begin{minipage}{0.28\textwidth}
    \includegraphics[width=\linewidth]{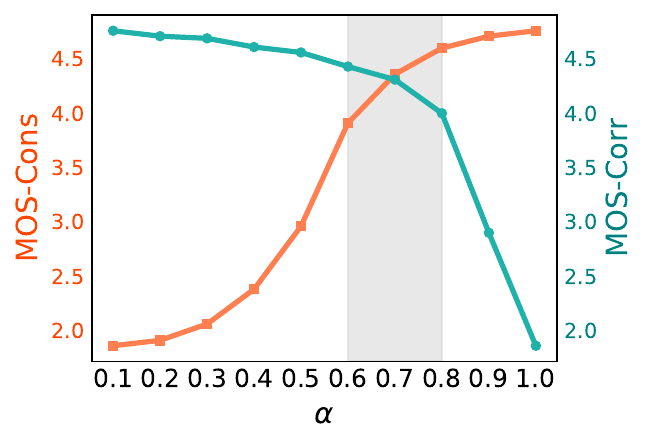}
  \end{minipage}%
  \begin{minipage}{0.20\textwidth}
    \captionsetup{width=\linewidth}
    \caption{MOS-Cons and MOS-Corr scores with varying editing degrees $\alpha$.  Edited speech tend to match both the source speech and text prompt in the highlighted area.}
    \label{fig_5}
  \end{minipage}
\vspace{-0.5cm}
\end{figure}

While the optimal value for editing degree $\alpha$ may vary depending on  different requirements, we aim to determine the optimal editing range for $\alpha$. The ideal voice attribute editing should involve a change toward the specified voice attribute direction while still preserving  some voice characteristics of the source speaker. To this end, we selected an additional  100 speech samples from unseen speakers and performed editing with weights $\alpha$ ranging from 0 to 1 across various attributes.  We evaluated the MOS-Cons and Mos-Corr scores of all generated speech and show the average results in Figure \ref{fig_5}. 

We observed that when the editing degree was less than 0.4, the generated speech closely resembled the source speech, and the editing had a small impact. For $\alpha \in [0.6, 0.8]$, the generated speech aligned well with the text prompt while also preserving the some voice characteristics of source speech. However, when the weight exceeded 0.8, there was a significant decrease in the similarity of voice characteristics between the generated speech and the source speech.  Therefore, we considered the optimal editing range to be between 0.6 and 0.8, setting the recommended value of $\alpha$ to 0.7.


\section{Conclusion}
In this work, we proposed \textit{VoxEditor}, the first voice attribute editing model with text prompt. Built upon an auto-encoder framework, we propose the ResMem and VADP blocks to effectively align voice attributes and the corresponding descriptors, facilitating quantifiable editing of speaker embeddings. Through experiments, we showcase the performance and generalization capability of \textit{VoxEditor} on both seen and unseen speakers. Experimental results  demonstrate that, with an appropriate editing degree, the generated speech not only meets the requirements of the text prompt but also retains the voice characteristics of source speech.  

\section*{Limitations} 
There are still limitations in data annotation and pre-training model aspects.
In terms of data annotation, we only annotated the prominent voice attributes during pairwise comparisons of the speaker's voice characteristics, while disregarding those diminished voice attributes. Those diminished voice attributes may be described by the text prompts such as "I hope this sound becomes less magnetic". If bidirectional annotations can be established, the model would also gain the capability for bidirectional voice attribute editing, thereby enhancing its overall performance.
Additionally, the number of annotated speakers in our dataset remains limited. Consequently, for descriptors with low frequency in the dataset, the corresponding voice attribute editing overly relies on a few specific speakers, thereby constraining the model's performance. In the future, we plan to explore automatic annotation models for voice characteristics, facilitating efficient dataset expansion. Furthermore, constrained by the zero-shot capability of the pre-trained VC network, VoxEditor exhibits slightly lower performance on the MOS-Cons and MOS-Corr metrics under unseen conditions compared to seen conditions. Therefore, scaling up our pretrained model will be our future work to further enhance performance.


\bibliography{custom}

\appendix

\section{Model Details}
\label{appendix A}

\subsection{Perturb-based Data Augmentation}
Following FreeVC, we distort the speaker information in the source waveform through three steps: (1) Obtain the original Mel spectrograms $mel_{ori}$ from the source speech waveform. (2) Apply spectrogram-resize (SR) to the original Mel spectrograms $mel_{ori}$ to obtain $mel_{sr}$. (3) Reconstruct the waveform from $mel_{sr}$. The SR process involves resizing the original Mel spectrograms along the frequency axis by a certain ratio $r$, followed by adjusting the resized Mel spectrograms to match the original size through padding or cutting. When $r > 1$, frequencies at the top are randomly removed, and when $r<1$, the top part of the Mel spectrogram is supplemented with Gaussian white noise atop the high-frequency portion.

\subsection{Speaker Encoder and Descriptor  Encoder}
To enhance the generalization of speaker embeddings, the speaker encoder in the model borrows the neural architecture from a speaker recognition network \cite{desplanques2020ecapa}, which is based on 1D-convolutional neural networks (1D-CNN) with an attentive statistics pooling layer.  The descriptor encoder comprises an embedding layer, a linear layer, and a ReLU activation function.

\subsection{Prior Encoder}
The prior encoder contains a bottleneck extractor and a normalizing flow. In order to remove the speaker information in self-supervised learning (SSL) feature extracted from the pretraind WavLM, we use a bottleneck extractor to map the 1024-dimension SSL feature into 192-dimension content embedding. The huge dimension gap is designed to discard non-essential content like noise or speaker information.  The linear projection layer above the blocks produces the mean and variance of the normal posterior distribution. The normalizing flow is used to improve the complexity of prior distribution. 

 \begin{table*}[!t]
\resizebox{\textwidth}{!}{
\begin{tabular}{lp{13.5cm}}
\toprule
\multicolumn{2}{c}{Diglogue}                                                     \\ \toprule
LLM prompts:    & Given a descriptor set for describing the common voice attribute: {[}Bright, Thin, Coarse, Delicate, Low, Pure, Rich, Magnetic, Muddy, Hoarse, Round, Flat, Shrill, Shriveled, Muffled, Soft, Transparent, Husky{]}, you will receive text prompt about relative modifications on specific voice attributes. Please search for the descriptors about the timbre attributes in the text prompt, and find the closest descriptors from the descriptor set as the answer, if there are more than one, please search in turn. Please reply in the following format: 1.voice attribute desciprot 1, 2. voice attribute descripor2. \\
Response: & Sure, please provide the text prompt for me to analyze and extract the   relevant voice attribute descriptors.   \\  \hdashline 
LLM prompts:    & I want the sound to be more magnetic  \\
Response: & From the provided text prompt "I want the sound to be more magnetic," the closest descriptor from the given set would be: 1. Magnetic\\  \hdashline                                  
LLM prompts:    & I want the sound to be more magnetic and bright \\                    Response: & From the provided text prompt "I want the sound to be more magnetic and bright," the closest descriptors from the given set would be: 1. Magnetic  2. Bright  \\ \hdashline 
LLM prompts:    & I want the sound to become more magnetic, bright and transparent  \\
Response: & From the provided text prompt   "I want the sound to become more magnetic, bright, and transparent," the closest descriptors from the given set would be: 1. Magnetic 2. Bright 3. Transparent   \\ \hdashline
LLM prompts:    & I want the voice to be hoarse and at the same time deeper  \\
Response: & From the provided text prompt   "I want the voice to be hoarse and at the same time deeper," the   closest descriptors from the given set would be: 1. Hoarse 2. Low   \\ \hdashline 
LLM prompts: & I want to achieve a lower-pitch sound. \\
Response: & From the provided text prompt   "I want to achieve a lower-pitch sound," the closest descriptor   from the given set would be: 1.Low  \\ \hdashline
LLM prompts:    & I'd like to add a touch of coarseness to this voice.  \\
Response: & From the provided text prompt   "I'd like to add a touch of coarseness to this voice," the closest   descriptor from the given set would be: 1. Coarse      \\ \bottomrule
\end{tabular}}
\caption{The examples of voice attributes extraction by the LLM.}
\label{table_3}
\end{table*}

The bottleneck extractor is consists of non-causal WaveNet residual blocks, containing layers of dilated convolutions with a gated activation unit and skip connection. The normalizing flow is a stack of affine coupling layers consisting of a  stack of WaveNet residual blocks.

\subsection{Posterior Encoder and Decoder}
In the posterior encoder, we employ non-causal WaveNet residual blocks following FreeVC \cite{li2023freevc}. The decoder essentially adopts the HiFi-GAN \cite{DBLP:conf/nips/KongKB20} generator structure, comprising stacked transposed convolutions, each followed by a multi-receptive field fusion module (MRF). The MRF's output is the aggregate of residual block outputs with varying receptive field sizes.

\subsection{Pretraining Strategy}
 \textit{VoxEditor} pipeline consists of two training stages. Firstly, we pretrain the FreeVC following the traditional VC task and then transfer the modules in FreeVC to \textit{VoxEditor}. During the training process 
 of \textit{VoxEditor}, we freeze the pretrained bottleneck extractor and speaker encoder to achieve speech representation disentanglement. 
 The networks were trained using the AdamW optimizer with $\beta_1=0.8$, $\beta_2=0.99$ and weight decay $\lambda=0.01$, with an initial learning rate of $2 \times 10^{-4}$.
 The pre-trained VC and VoxEditor were both trained on a single NVIDIA 4090 GPU with a batch size of 64 and a maximum segment length of 128 frames,
for 900k steps and 150k steps, respectively.

\subsection{Training Loss}
In general, the training loss of \textit{VoxEditor} is divided into CVAE-related loss $\mathcal{L}_{cvae}$, Text-Voice alignment-related loss  and GAN-related loss $\mathcal{L}_{gan}$. 
The $\mathcal{L}_{cvae}$ and $\mathcal{L}_{gan}$ follows autoencoder training loss functions of FreeVC. The GAN-related loss consists of adversarial loss  $\mathcal{L}_{adv}(D)$ and $\mathcal{L}_{adv}(G)$ for discriminator $D$ and generator $G$ and feature matching loss $\mathcal{L}_{fm}(G)$ for generator $G$.
The Text-Voice alignment-related loss contains  $\mathcal{L}_{rec}$ in Equation \ref{eq7} and ${\mathcal{L}_{align}}$ in Equation \ref{eq12}. The final loss function during the training process of \textit{VoxEditor} is as follows,
\begin{multline} \label{eq14}
\mathcal{L(G)}=\mathcal{L}_{cvae} + \lambda _ 1 \mathcal{L}_{rec} + \lambda _ 2  \mathcal{L}_{align}  \\ 
+  \mathcal{L}_{adv}(G) + \mathcal{L}_{fm}(G),
\end{multline}
\begin{equation}
  \mathcal{L(D)}=\mathcal{L}_{adv}(D), 
\end{equation}
where ${\lambda_1}$ and ${\lambda_2}$ in Equation \ref{eq14} were respectively set to be 20 and 200.

\begin{table}[]
\begin{tabular}{cccccc}
\toprule
\multirow{2}{*}{M} & \multirow{2}{*}{N} & \multicolumn{2}{c}{Seen} & \multicolumn{2}{c}{Unseen} \\ \cmidrule(r){3-4} \cmidrule(l){5-6} 
                   &                    & TVAS        & SS         & TVAS         & SS          \\ \midrule
16                 & 4                  & 0.0546      & 0.8361     & 0.0537       & 0.7012      \\
32                 & 4                  & \textbf{0.0574}      & 0.8830     & \textbf{0.0561}       & \textbf{0.7252}      \\
64                 & 8                  & 0.0561      & 0.8428     & 0.0554       & 0.7202      \\
96                 & 16                 & 0.0568      & \textbf{0.8831}     & 0.0559       & 0.7251      \\ \bottomrule
\end{tabular}
\caption{The  performance evaluation for the \textit{VoxEditor} with different hyperparameters for the ResMem block.}
\label{table4}
\end{table}

\section{LLM Prompts for Descriptor Extraction from Text Prompt}
\label{appendix B}
Table \ref{table_3} provides a detailed example of voice attribute extraction, illustrating the LLM prompts and responses.

\section{Hyperparameter Selection for the ResMem Block}
\label{App ResMem block}
In this section, we provide a detailed explanation of the hyperparameter selection for the ResMem Block introduced in Section \ref{ResMem Block}. The ResMem block primarily comprises two key hyperparameters: the slot number $M$ in the main voice-value memory $\bm{M}_{mv}$ and slot number $N$ in the residual voice-value memory $\bm{M}_{rv}$. We trained the \textit{VoxEditor} using various combinations of  $M$ and $N$, evaluating the TVAS and speaker similarity between source speech with reconstructed speech (SS). Here, the reconstructed speech refers to the edited speech with $\alpha=0$. As depicted in Table \ref{table4}, the optimal overall performance of the \textit{VoxEditor} is achieved when $M=32$ and $N=4$.

\section{Ablation Models}
\label{Ablation Models}
We extensively discussed the configuration of the ablation methods in Section 
\ref{Evaluation Results} as follows.
\paragraph{w/o Voice Res.}: Voice-value residual memory in the ResMem block was removed, making recalled speaker embeddings equal to the recalled main speaker embeddings. The original Equation \ref{eq7} and Equation \ref{eq13} were  transformed as follows,
\begin{equation}
\mathcal{L}_{rec}=\| \bm{s} - \hat{\bm{s}} \Vert _2 ^2.
\label{eq16}
\end{equation}
\begin{equation}
{\bm{s}_{edit}} = \alpha \cdot \hat{\bm{t}} + (1-\alpha) \cdot \hat{\bm{s}}_{m}.
\end{equation}

\paragraph{w/o ResMem}: the ResMem block was removed and the output of speaker encoder and descriptor encoder was directly interpolate.  
\paragraph{w/o VADP}: the VADP block was removed. The original Equation \ref{eq11} and Equation \ref{eq12} were transformed as follows,
 \begin{equation}
{\bm{s}_{edit}} =  \hat{\bm{t}} + \hat{\bm{s}}_{m}^A,
\end{equation}
\begin{equation} 
\mathcal{L}_{align}=D_{KL}(\bm{w}_{mv}^{B}||\bm{w}_t + \bm{w}_{mv}^{A}).
\end{equation}
\end{document}